\documentclass[aps,prd,twocolumn,superscriptaddress]{revtex4-2}

\usepackage[dvips]{graphicx}
\usepackage{ulem}
\usepackage[usenames]{color}
\usepackage{amsmath}
\usepackage{amssymb}
\usepackage{bm}
\usepackage{mathrsfs}
\usepackage{amsbsy}
\usepackage{hyperref}
\usepackage[all]{hypcap}
\usepackage{longtable}
\usepackage{url}
\usepackage{multirow}

\begin{document}

\title{$X(3872)$ and hidden charmed tetraquarks}

\author{You-You Lin}

\author{Ji-Ying Wang}

\author{Ailin Zhang}
\email{zhangal@shu.edu.cn}
\affiliation{Department of Physics, Shanghai University, Shanghai 200444, China}

\begin{abstract}
In a constituent quark model, a hidden charmed tetraquark is assumed consisting of a $cq$ diquark and an $\bar c\bar q$ antidiquark or vice versa. The Semay-Silvestre-Brac potentials are employed to calculate the masses of $cq$ (q=u, d) diquarks. The mass of the $cq$ diquark or $\bar c\bar q$ antidiquark with spin-$0$ is predicted with $\sim 2175$ MeV, and the spin-$1$ one is predicted with $\sim 2220$ MeV. The masses of hidden charmed tetraquarks from $1S$ to $2P$ excitations are systematically calculated in terms of the same potentials. It is found that the mass of hidden charmed tetraquarks without radial excitation grows higher in $1^{+-},~1^{++},~1^{--},~0^{-+},~0^{--},~1^{-+},~\cdots$ sequence, and the tetraquarks with exotic $J^{PC}=0^{--},~1^{-+}$ have higher masses. The hidden charmed tetraquarks with radial excitations have masses greater than $4300$ MeV. The $1S-1P$ and $1S-2S$ mass splittings of the hidden charmed tetraquarks are about $390-400$ MeV and $550-570$ MeV, respectively, which are about $50$ and $40$ MeV smaller than those of normal charmoniums. The $1P-2P$ and $2S-2P$ mass splittings are similar to those for conventional $c\bar c$ charmonium mesons. Based on our predicted masses for hidden charmed tetraquarks, some XYZ exotics are analyzed and tentatively assigned. $X^*(3860)$ is possibly the $0^{++}$ tetraquark. $Z_c(3900)$ and $X(3940)$ are possibly the $1^{+-}$ tetraquarks, and $X(3872)$ is possibly a $1^{++}$ tetraquark. $X(4250)$ may be a $0^{-+}$, $0^{++}$ or $1^{-+}$ tetraquark, $X(4240)$ may be a $0^{--}$ tetraquark. With radial excitations, $X(4350)$ may be a $0^{++}$ tetraquark, $Z_c(4430)$ may be a $1^{+-}$ tetraquark, $X(4630)$ may be a $0^{-+}$ or $1^{-+}$ tetraquark, and $X(Y)(4660)$ may be the $1^{--}$ tetraquark. $Y(4008)$ or $Y(4390)$ seems impossible to be the $1^{--}$ tetraquark. In experiments, for the identification of a normal charmonium or a charmonium-like exotic, it is also important to take into account the mixing of normal charmoniums with hidden-flavor charmonium-like exotics.
\end{abstract}

\maketitle
\section{Introduction}\label{intro}
A four-quark state consists of two quarks and two antiquarks. The concept of four-quark states was first mentioned by Gell-Mann, and four-quark states were subsequently studied in hadron scattering amplitudes. The discovery of $X(3872)$ stirred great interest in four-quark states. $X(3872)$ was first observed in exclusive decays $B^\pm$ with $X(3872) \to \pi^+\pi^- J/\Psi$ by Belle~\cite{Belle:2003nnu}, where $\pi^+\pi^-$ is found to be
dominated by $\rho^0$ with isospin symmetry violating. Later, the isospin violating channel $X(3872)\to \rho^0 J/\psi $~\cite{LHCb:2022jez} was observed. A surprising $\pi^+\pi^-\pi^0 J/\Psi$ mode was also reported by Belle~\cite{Belle:2005lfc}, but subsequent experiments by Belle~\cite{Belle:2010sgx,Belle:2011jbs} did not observe such a $3\pi J/\psi$ decay mode in the radiative decays of $\Upsilon(1S)$ and $\Upsilon(2S)$. However, the isospin conserving channel $X(3872)\to \omega J/\psi $ was observed~\cite{Belle:2005lfc,BaBar:2010wfc,BESIII:2019qvy}. The isospin, G parity and $J^{PC}$ quantum numbers of $X(3872)$ have finally been determined with $I^G(J^{PC})=0^+(1^{++})$~\cite{ParticleDataGroup:2024cfk}.

In addition to $X(3872)$, many hidden charmed tetraquark states have been observed by different experimental collaborations. Hidden charmed states are states that contain both a $c$ and a $\bar c$ component. In Table~\ref{candidates}, the hidden charmed exotics known as XYZ states without strange or antistrange quark component in the present Particle Data Group (PDG)~\cite{ParticleDataGroup:2024cfk} are listed, where the measured masses and widths are presented. The quantum numbers $J$ and $P$ of $X(3823)$ need confirmation. The new naming scheme for the neutral-flavor hidden charm exotics is employed in the second column. Some of the exotics have been renamed as normal charmoniums in the present PDG though these assignments are not definite. The last column indicates the experimental collaborations to the first relevant observations.
\begin{table*}[htb]
\caption{\label{candidates}Observed hidden charmed states without $s$ or $\bar s$ quark~\cite{ParticleDataGroup:2024cfk}. The last column indicates the first observation.}
\begin{ruledtabular}
\begin{tabular}{ccccccc}
$J^{PC}$ & States & $M_{exp}$ (MeV) & $\Gamma$ (MeV) & Known as & Experiments \\
\colrule
$2^{--}$, J, P? & $\Psi_2(3823)$   & $3823.51\pm0.34$ & $<2.9$ & $\Psi(3823),~X(3823)$ & Belle  \\
$0^{++}$ & $\chi_{c0}(3860)$   & $3862^{+26+40}_{-32-13}$ & $201^{+154+88}_{-67-82}$ & $X^*(3860)$ & Belle  \\
$1^{++}$ & $\chi_{c1}(3872)$ & $3871.64\pm0.06$  & $1.19\pm0.21$  & X(3872) & Belle \\
$1^{+-}$ & $T_{c\bar c1}(3900)$   & $3887.1\pm2.6$  & $28.4\pm2.6$  & $Z_c(3900)$ & BESIII\\
$0^{++}$ & $\chi_{c0}(3915)$ & $3922.1\pm1.8$  & $20\pm6$  & X(3915) & Belle \\
$?^{??}$ & $X(3940)$   & $3942\pm9$ & $43^{+28}_{-18}$ & $X(3940)$ & Belle  \\
$?^{?-}$ & $T_{c\bar c1}(4020)$   & $4024.1\pm1.9$  & $13\pm5$  & $Z_c(4020)$ & BESIII\\
$?^{?+}$ & $T_{c\bar c1}(4050)$   & $4051\pm14^{+20}_{-41}$  & $45\pm11\pm6$  & $X(4050)$ & Belle\\
$?^{?-}$ & $T_{c\bar c1}(4055)$   & $4054\pm3\pm1$  & $82^{+21+47}_{-17-22}$  & $X(4055)$ & Belle\\
$?^{?+}$ & $T_{c\bar c1}(4100)$   & $4096\pm20^{+18}_{-22}$  & $152\pm58^{+60}_{-35}$  & $X(4100)$ & LHCb\\
$1^{++}$ & $\chi_{c1}(4140)$ & $4146.5\pm3.0$  & $19^{+7}_{-5}$  & X(4140) & CDF \\
$?^{??}$ & $X(4160)$ & $4153^{+23}_{-21}$  & $136^{+60}_{-35}$  & X(4160) & Belle \\
$1^{+-}$ & $T_{c\bar c1}(4200)$   & $4196^{+31+17}_{-29-13}$  & $370\pm70^{+70}_{-132}$  & $Z_c(4200),~X(4200)$ & Belle\\
$1^{--}$ & $\Psi(4230)$   & $4222.2\pm2.4$  & $51\pm8$  & $Y(4230),~Y(4260)$ & BABAR\\
$0^{--}$ & $T_{c\bar c0}(4240)$   & $4239\pm18^{+45}_{-10}$  & $220\pm47^{+108}_{-74}$  & $Z_c(4240),~X(4240)$ & LHCb\\
$?^{?+}$ & $T_{c\bar c}(4250)$   & $4248^{+44+180}_{-29-35}$  & $177^{+54+316}_{-39-61}$  & $X(4250)$ & Belle\\
$1^{++}$ & $\chi_{c1}(4274)$ & $4286^{+8}_{-9}$  & $51\pm7$  & X(4274) & CDF,~LHCb \\
$?^{?+}$ & $X(4350)$   & $4350.6^{+4.6}_{-5.1}\pm0.7$  & $13^{+18}_{-9}\pm4$  & $X(4350)$ & Belle\\
$1^{--}$ & $\Psi(4360)$   & $4374\pm7$  & $120\pm12$  & $X(4360),~Y(4360)$ & Belle\\
$1^{--}$ & $Y(4390)$   & $4382.0\pm13.3\pm1.7$  & $135.8\pm60.8\pm22.5$  & $Y(4390)$ & BESIII~\cite{BESIII:2020bgb}\\
$1^{+-}$ & $T_{c\bar c1}(4430)$   & $4478^{+15}_{-18}$  & $181\pm31$  & $Z_c(4430),~X(4430)$ & Belle\\
$0^{++}$ & $\chi_{c0}(4500)$ & $4474\pm3\pm3$  & $77\pm6^{+10}_{-8}$  & X(4500) & LHCb \\
$?^{?+}$ & $X(4630)$   & $4626\pm16^{+18}_{-110}$  & $174\pm27^{+134}_{-73}$  & $X(4630)$ & LHCb\\
$1^{--}$ & $\Psi(4660)$   & $4623\pm10$  & $55\pm9$  & $X(4660),~Y(4660)$ & Belle\\
$0^{++}$ & $\chi_{c0}(4700)$ & $4694\pm4^{+16}_{-3}$  & $87\pm8^{+16}_{-6}$  & X(4700) & LHCb
\end{tabular}
\end{ruledtabular}
\end{table*}

For its exotic properties that are different from a conventional $c\bar c$ meson, $X(3872)$ has been extensively studied and assigned in all possible models. Many experiments have also been performed for its different features in different facilities~\cite{CDF:2003cab,BaBar:2004oro,CDF:2006ocq,BaBar:2008flx,Belle:2011wdj,CMS:2013fpt,LHCb:2013kgk,BESIII:2013fnz,LHCb:2015jfc,LHCb:2020xds}.
So far, there are many interpretations of $X(3872)$ based on different concerns from different perspectives. $X(3872)$ was assigned with a $(c\bar q)(q\bar c)$-like hadronic molecule in Refs.~\cite{Close:2003sg,Voloshin:2003nt,Tornqvist:2004qy,Voloshin:2004mh,Braaten:2003he,Wong:2003xk,Hogaasen:2005jv,Braaten:2007dw,Ortega:2009hj,Liu:2019stu,Liu:2019tjn}, and with a tetraquark state consisting of a $cq$ diquark and an $\bar c\bar q$ antidiquark in Refs.~\cite{Maiani:2004vq,Ebert:2005nc,Terasaki:2007uv,Dubnicka:2010kz,Kleiv:2013dta,Maiani:2014aja,Zhao:2014qva,Anwar:2018sol,Grinstein:2024rcu}. It was interpreted as a hybrid charmonium in Ref.~\cite{Li:2004sta}, and as an adjoint charmonium in Ref.~\cite{Braaten:2013boa}. $X(3872)$  was interpreted as a normal $\chi_{c1}(2P)$ charmonium in Ref.~\cite{Barnes:2003vb}, and as a mixing state, for example, a normal $c\bar c$ charmonium mixing with $D\bar D^*+\bar DD^*$~\cite{Suzuki:2005ha} or diquark-antidiquark mixing with a molecule~\cite{Yang:2009zzp}. It was also interpreted as kinematic or threshold effects from virtual particles~\cite{Kalashnikova:2005ui,Kalashnikova:2009gt,Danilkin:2010cc}. Among these different interpretations, the hadronic molecule interpretation of $X(3872)$ was the one most popularly advocated.

A diquark was first mentioned by Gell-Mann~\cite{GellMann1964}, and subsequently introduced to describe a baryon as a composite state of two bodies, a quark and a diquark~\cite{Ida:1966ev,Lichtenberg:1967zz,Fleck:1988vm,Selem:2006nd,Chen:2009tm,Ferretti:2011zz,Ebert:2011kk,Chen:2014nyo,Kumakawa:2017ffl}. In this way, the puzzle of supernumerary baryons could be understood and some features of baryon spectroscopy were well described. Later, the diquark as a fundamental composite was widely employed to construct many kinds of multiquark systems~\cite{Rosenzweig:1975fq,Anselmino:1992vg,Jaffe:2003sg,Karliner:2003dt,Maiani:2004uc,Maiani:2004vq,Jaffe:2004ph,Maiani:2015vwa,Chen:2016qju,Lebed:2016hpi,Maiani:2017kyi,Olsen:2017bmm,Barabanov:2020jvn}.
The diquark cluster in hadrons is in fact a kind of strong correlation between a pair of quarks. The two quarks attract each other and make up a totally antisymmetric diquark, where the two quarks are also in an antisymmetric color representation. There are two kinds of antisymmetric light diquarks, scalar diquark and vector diquark. The scalar diquark with spin-$0$ has a lower mass in comparison to the vector diquark with spin-$1$. They are usually called the ``good" and ``bad" diquarks, respectively, as suggested in Ref.~\cite{Jaffe:2004ph}. The diquark is thought either as stuff without internal structure or as a cluster with a finite size.

Phenomenological evidence for diquarks is mainly found in baryon spectroscopy, the $\Delta I={1\over 2}$ rule in weak nonleptonic decays, and some features of parton distribution functions~\cite{Jaffe:2003sg,Jaffe:2004ph}. Recently, a positive measure of the production cross sections of hyperons and charmed baryons from $e^+e^-$ annihilation was performed by the Belle collaboration~\cite{Belle:2017caf}. However, there is not yet any conclusive experimental evidence for the existence of diquark cluster in baryons~\cite{Sateesh:1991jt,Leinweber:1993nr,Glozman:1999pt}, and it is not clear either which system is the right one for two quarks to make a diquark.

From a phenomenological analysis~\cite{Jaffe:2004ph}, the mass difference between good and bad light diquarks is estimated with $\sim 200$ MeV (${2\over 3}$ of the $\Delta$-nucleon mass difference) in light baryons~\cite{Jaffe:2004ph}. The mass difference is about $212$ MeV in baryons with one heavy quark where the mass difference between a good light diquark and a light quark is $\sim 310$ MeV. Masses of diquarks have also been calculated in many different methods such as the Bethe-Salpeter equation~\cite{Cahill:1987qr,Bender:1996bb,Maris:2002yu,Roberts:2011cf}, constituent quark model~\cite{Ebert:2002ig,Maiani:2004vq,Ebert:2005xj,Ebert:2011kk,Ferretti:2011zz,Santopinto:2014opa,Kumakawa:2017ffl,Lin:2024gcm}, Faddeev equations~\cite{Hecht:2002ej}, instanton model~\cite{Schafer:1993ra}, QCD sum rules~\cite{Zhang:2006xp,Kleiv:2013dta,Esau:2019hqw,deOliveira:2023hma}, and lattice theory~\cite{Watanabe:2021nwe,Francis:2021vrr}.

There were also some investigations on how to distinguish different interpretations. With a study of the line shape of $X(3872)$, the hadronic molecule interpretation may be distinguished from the virtual charm mesons interpretation~\cite{Braaten:2007ft}. Through the mass spectrum, different patterns for the XYZ states in the hadro-charmonium picture, the tetraquark picture as well as the hadronic molecular picture have been explored~\cite{Cleven:2015era}. Through an effective range defined through the low energy expansion of the scattering phase, the compact tetraquark nature and the loosely bound molecular one to a state can be discriminated~\cite{Esposito:2021vhu}. In the Born-Oppenheimer effective theory, the authors in Ref.~\cite{Berwein:2024ztx} claimed that XYZ exotics of any composition with two heavy quarks such as hybrids, tetraquarks, pentaquarks, doubly heavy baryons, and quarkonia can all be addressed. Polosa {\it et al.} suggested that a measure of the ratio $R=\frac{B(X\to\psi'\gamma)}{ B(X\to\psi\gamma)}$ for radiative decays of $X(3872)$ can serve as a strong indication to discriminate between the molecular and compact interpretations~\cite{Grinstein:2024rcu}. In their calculation, $R\sim 1$ or larger was found for a compact diquark-antidiquark state and $R$ was of order $10^{-2}$ for a molecule state. In $pp$ collisions, some observables were proposed to serve as criteria to identify the $X(3872)$ as either a molecular state or a compact tetraquark state~\cite{She:2024cit}.

From the $pp$ collision data, the $X(3872)\to \psi(2S)\gamma$ process is observed for the first time by the LHCb collaboration. In particular, the ratio of two partial widths is measured with
\begin{equation*}
R=\frac{\Gamma_{X(3872)\to\psi(2S)\gamma}}{ \Gamma_{X(3872)\to\psi\gamma}}=1.67\pm 0.21\pm 0.12 \pm 0.04.
\end{equation*}
Based on the proposal and calculation by Polosa et al.~\cite{Grinstein:2024rcu}, the measured ratio does not support the pure molecule interpretation and strongly indicates a sizeable compact charmonium or tetraquark component within $X(3872)$.

In the theoretical point of view, the interpretations of exotic states are based on either phenomenological investigations or on effective quantum field theory. In fact, the interpretation of $X(3872)$ or other exotic states depends on the interactions and models employed. How to identify $X(3872)$ among different interpretations is always a puzzle and problem waiting to be solved. So far, there is not a single theoretical model to give a unified understanding of the spectroscopy, decay patterns, production features, and structure of those exotic states. However, the features of many XYZ exotics are still unknown, and the quantum numbers $J$, $P$ and $C$ (if they exist) of some exotics are not determined.

According to current wisdom, quantum chromodynamics (QCD) is the proper theory to describe strong interaction, and every system confined through strong interaction in principle could be understood from QCD or lattice QCD. However, with the limitations of computers at present, many topics related to nonperturbation and confinement have to rely on effective theories or phenomenological theories. Constituent quark potential models have made great progress in the understanding of conventional mesons and baryons, and have been employed to study exotic hadrons.

In constituent quark potential models, the spectra of different tetraquarks have been calculated in a diquark-antidiquark picture in many studies, and more possible calculations of hidden charmed tetraquarks are expected. The spectrum of $c\bar q c\bar q$ tetraquarks was calculated by means of an algebraic mass formula where the mass of $X(3872)$ was set as an input~\cite{Maiani:2004vq}. The tetraquark spectrum was calculated in a relativistic diquark-antidiquark model with one-gluon exchange and long-range vector and scalar linear confining potentials, where the diquarks are considered to have an internal structure with form factor~\cite{Ebert:2007rn}. The spectra of hidden charmed $[qc][\bar q\bar c]$ and $[sc][\bar s\bar c]$ tetraquarks were calculated in a relativized diquark model where the diquark masses are extracted from  previous studies~\cite{Anwar:2018sol}.

In this paper, a hidden charmed tetraquark without a strange or antistrange quark component is thought as a composite with a compact $cq$ (q=u, d) diquark and a compact $\bar c\bar q$ antidiquark. The interactions between a quark and a quark, an antiquark and an antiquark, a quark and an antiquark, and a diquark and an antidiquark are described by universal Semay-Silvestre-Brac potentials~\cite{Semay:1994ht,Silvestre-Brac:1996myf}. These potentials have been recently employed to calculate the masses of the doubly charmed tetraquarks~\cite{Lin:2024gcm}. In the practical calculation process, the masses of relevant diquarks are first computed in terms of the potentials, and the masses of hidden charmed tetraquarks from $1S$ to $2P$ excitations are subsequently computed. In this way, some features of the mass spectra are obtained for hidden charmed tetraquarks with different inner structures. Based on our numerical results, all the observed exotics in Table~\ref{candidates} are examined, and possible assignments of some XYZ exotics are made.

The work is organized as follows, a simple description of the constituent quark potential model is presented in Sec.~\ref{model} after the introduction. Section~\ref{results} is devoted to the practical calculation and analyses of the observed XYZ exotics. The last section gives a brief summary and discussion.

\section{Constituent quark potential model}\label{model}
For a practical calculation, a modified potential for the quark-antiquark interaction in a norelativistic potential model proposed by Semay and Silvestre-Brac ~\cite{Semay:1994ht,Silvestre-Brac:1996myf} was employed, with quark-antiquark interaction potential
\begin{equation}\label{eq:brac}
\begin{aligned}
V_{q\bar q}=&V_0+(\vec{s}_q \cdot \vec{s}_{\bar q})V_{ss}\\
=&-\frac{\alpha(1-e^{-r/r_c})}{r}+\lambda r^{p}+C\\
&+(\vec{s}_q \cdot \vec{s}_{\bar q})\frac{8\kappa(1-e^{-r/r_c})}{3m_q m_{\bar q}\sqrt{\pi}}\frac{e^{-r^2/r_0^2}}{r_0^3}
\end{aligned}
\end{equation}
with
\begin{equation*}\label{eq:r0}
	r_0=A{\left(\frac{2m_qm_{\bar q}}{m_q+m_{\bar q}}\right)}^{-B}.
\end{equation*}
The details on the parameters can be found in Refs.~\cite{Semay:1994ht,Silvestre-Brac:1996myf} in which the parameters are determined through a systematic investigation on mesons and baryons. The quark-quark interaction potential in baryon or multiquark states is similar to $V_{qq}={\frac{1}{2}}V_{q\bar q}$.


The full Hamiltonian of hidden charmed tetraquarks is constructed in a similar way as that in Ref.~\cite{Lin:2024gcm},
\begin{equation}\label{eq:ham}
H=m_d+m_{\bar d}+\frac{p^2}{2\mu}+V_{al}+V_{sl},
\end{equation}
where $V_{al}$ is the $AL$-type potential~\cite{Silvestre-Brac:1996myf} consisting of one gluon exchange interaction and a linear confinement term, and $V_{sl}$ is a universal Breit-Fermi approximation responsible for the spin and orbital angular momentum coupling~\cite{Lin:2024gcm,DeRujula:1975qlm}
\begin{equation}\label{eq:vsl}
\begin{aligned}
V_{sl}=&V_{so}+V_{ten}=\frac{\alpha(1-e^{-r/r_c})}{r^3}\left(\frac{1}{m_d}+\frac{1}{m_{\bar d}}\right)\\
&\times\left(\frac{\vec{s}_d}{m_d}+\frac{\vec{s}_{\bar d}}{m_{\bar d}}\right) \cdot \vec{L}-\frac{1}{2r}\frac{\partial V_{conf}}{\partial r}\left(\frac{\vec{s}_d}{m_d^2}+\frac{\vec{s}_{\bar d}}{m_{\bar d}^2}\right) \cdot \vec{L}\\
&+\frac{1}{3m_d m_{\bar d}}\left(\frac{1}{r}\frac{\partial V_{Coul}}{\partial r}-\frac{\partial^2V_{Coul}}{\partial r^2}\right)\\
&\times\left(\frac{3\vec{s}_d \cdot \vec{r} \vec{s}_{\bar d}  \cdot \vec{r}}{r^2}-\vec{s}_d \cdot \vec{s}_{\bar d}\right).	
\end{aligned}
\end{equation}

For the calculation of masses of the $cq$ diquark and $\bar c\bar q$ antidiquark, the Hamiltonian is the same as Eq. (\ref{eq:ham}) except that the following quark-quark potential inside the diquark/antidiquark was employed
\begin{equation}\label{eq:vcc}
\begin{aligned}
V_{cq}=&\frac{1}{2}\left[-\frac{\alpha(1-e^{-r/r_c})}{r}+\lambda r+C\right.\\
&\left.+(\vec{s}_c \cdot \vec{s}_q)\frac{8\kappa(1-e^{-r/r_c})}{3m_c m_q\sqrt{\pi}}\frac{e^{-r^2/r_0^2}}{r_0^3}\right].
\end{aligned}
\end{equation}

The total wave function of a hidden charmed tetraquark in the diquark-antidiquark picture
\begin{equation}
	\Psi_{JM}=\chi_c\otimes\chi_f\otimes[\Psi_{lm}(\vec{r})\otimes \chi_s]_{JM},
\end{equation}
is composed of color, flavor, spin and spatial wave functions $\chi_c$, $\chi_f$, $\chi_s$ and $\Psi_{lm}(\vec{r})$. Due to the interaction in diquarks, only color $\mathbf{\bar{3}}_c(\mathbf{3}_c)$ diquark (antidiquark) is considered in this work. Therefore two kinds of diquarks/antidiquarks satisfy the symmetry constraints: the flavor symmetric one with spin-$1$ denoted as $\{cq\}^{\bar 3}_1$/$\{\bar{c}\bar{q}\}^3_1$ and the flavor antisymmetric one with spin-$0$ denoted as $[cq]^{\bar 3}_0$/$[\bar{c}\bar{q}]^3_0$. In this way, the constructed wave functions and corresponding quantum numbers of some ground state tetraquarks are listed~\cite{Maiani:2004vq,Giron:2019cfc,Anwar:2018sol}
\begin{flalign}\label{eq:cs}
     J^{PC}=0^{++}:
     & \chi_1=|[cq]^{\bar 3}_0[\bar{c}\bar{q}]^3_0\rangle_0,\notag\\
     & \chi_2=|\{cq\}^{\bar 3}_1 \{\bar{c}\bar{q}\}^3_1\rangle_0,\notag\\
	 J^{PC}=1^{++}:
     & \chi_3=\frac{1}{\sqrt{2}}|\{cq\}^{\bar 3}_1[\bar{c}\bar{q}]^3_0+[cq]^{\bar 3}_0\{\bar{c}\bar{q}\}^3_1\rangle_1,\notag\\
     J^{PC}=1^{+-}:
     & \chi_4=\frac{1}{\sqrt{2}}|\{cq\}^{\bar 3}_1[\bar{c}\bar{q}]^3_0-[cq]^{\bar 3}_0\{\bar{c}\bar{q}\}^3_1\rangle_1,\notag\\
     & \chi_5=|\{cq\}^{\bar 3}_1 \{\bar{c}\bar{q}\}^3_1\rangle_1,\notag\\
	 J^{PC}=2^{++}:
     & \chi_6=|\{cq\}^{\bar 3}_1 \{\bar{c}\bar{q}\}^3_1\rangle_2,
\end{flalign}
where $P=(-1)^L$ with $L$ indicating the orbital angular momentum between the diquark and the antidiquark. For neutral-flavor hidden charmed tetraquarks, they may have definite $C$ quantum numbers. On a convenient basis for the $C$ parity, the $C$ parity of a tetraquark is defined in Ref.~\cite{Anwar:2018sol} after recoupling of the quark spins
\begin{align}
C&=(-1)^{L+s_{c\bar c}+s_{q\bar q}},
\end{align}
where $s_{c\bar c}$ is the spin of the charm-anticharm quarks, and $s_{q\bar q}$ is the spin of the light quark-antiquark~\cite{Anwar:2018sol}. The tetraquark state will be denoted as $n|[s_d,s_{\bar d}]_S,L\rangle_{J}$ with the radial quantum number $n$, total spin $S$ and total angular momentum $\vec J=\vec L+\vec S$.

For normal neutral-flavor quark-antiquark mesons, possible $J^{PC}$ quantum numbers are $P=(-1)^{L+1}$ and $C=(-1)^{L+S}$. Accordingly, quarkonia has natural $J^{PC}$ quantum numbers such as $0^{-+}$, $0^{++}$, $1^{--}$, $1^{+-}$, $1^{++}$, $2^{--}$, $2^{-+}$, $2^{++}$, $\cdots$, and cannot have unnatural $J^{PC}$ quantum numbers such as $0^{--}$ and $1^{-+}$. For neutral-flavor tetraquarks, they may have such exotic quantum numbers that are unnatural for quarkonia.

In order to solve the Schr\"{o}dinger equation, the spatial wave function $\psi_{lm}(\vec{r})$ is expanded in terms of Gaussian functions as the trial wave function as done in Ref.~\cite{Hiyama:2003cu}. Therefore, the Schr\"{o}dinger equation is transformed into a matrix eigenvalue problem which can be solved as the procedure in Refs.~\cite{Hiyama:2003cu,Lin:2024gcm}. The methods in Ref.~\cite{Lin:2024gcm} developed for doubly charmed tetraquarks are now applied to tetraquarks with hidden charm.

\section{Diquarks and hidden charmed tetraquarks}\label{results}
To proceed to the calculation, the constituent quark masses and the parameters in the $AL$ quark-quark potentials for the $cq$ (q=u, d) diquark and antidiquark are employed as those in Refs.~\cite{Semay:1994ht,Silvestre-Brac:1996myf,Lin:2024gcm}.

In terms of those parameters in Refs.~\cite{Semay:1994ht,Silvestre-Brac:1996myf}, the masses of $cq$ diquarks and antidiquarks are computed with the $AL$ potentials, and the results are presented in Table~\ref{tab:mdd}. As a comparison, the corresponding masses from a phenomenological analysis~\cite{Maiani:2004vq} and a QCD sum rule calculation~\cite{Kleiv:2013dta} are also presented.
\begin{table}[htb]
\caption{\label{tab:mdd}Mass (in MeV) of the $cq$ (q=u, d) diquark/antidiquark in $AL$ potentials.}
\begin{ruledtabular}
\begin{tabular}{cccccc}
spin & $AL1$  & $AL2$  & Ref.\cite{Maiani:2004vq} & Ref.~\cite{Kleiv:2013dta}\\
\colrule
$s_d=0$ & $2169.63$ & $2184.92$ & $1933$ & $1860$\\
$s_d=1$ & $2211.89$ & $2227.64$ & $1933$ & $1870$
\end{tabular}
\end{ruledtabular}
\end{table}

The $cq$ diquark with spin-$1$ has a mass $\sim43$ MeV greater than that of the $cq$ diquark with spin-$0$. The predicted mass of the $cq$ diquark with spin-$1$ is about $\sim 300$ MeV greater than the corresponding one from the phenomenological analysis~\cite{Maiani:2004vq} or the QCD sum rule~\cite{Kleiv:2013dta}.

In practical calculations of spectra of hidden charmed tetraquark states, the two parameters $\alpha$ and $\lambda$ employed in this work are slightly different from those for conventional mesons and baryons. According to the argument in Ref.~\cite{Lin:2024gcm}, the parameters $\alpha$ and $\lambda$ in the $AL$ quark-antiquark potential between a $cq$ diquark and a $\bar c\bar q$ antidiquark are fixed through two ground state tetraquark candidates: $X(3872)$ and $T_{cc}(3875)^+$. $T_{cc}(3875)^+$ is assumed as a doubly charmed $J^P=1^+$ tetraquark made of a $cc$ diquark and a $\bar u\bar d$ antidiquark~\cite{Lin:2024gcm}, and $X(3872)$ is assumed as a hidden charmed $J^{PC}=1^{++}$ ground state tetraquark made from a $cq$ diquark and an $\bar c\bar q$ antidiquark. As indicated in Ref.~\cite{Lin:2024gcm}, the real components of these two states (including mixing) will bring in a small uncertainty to the parameters $\alpha$ and $\lambda$. A combination of two types of AL potentials applied to diquarks/antidiquarks leads to four sets of parameters. The four sets of $\alpha$ and $\lambda$ are employed as those in Ref.~\cite{Lin:2024gcm}. The spin-spin interaction does not contribute to the masses of these two states, and the values of $\kappa$, $C$, $B$, $A$ and $r_c$ follow the original sets in Refs.~\cite{Semay:1994ht,Silvestre-Brac:1996myf}.

In terms of the parameters shown above and the obtained masses of diquarks/antidiquarks, the mass spectra of hidden charmed tetraquarks from $1S$ to $2P$ excitations are calculated through the solution of matrix eigenvalue problem as done in Refs.~\cite{Hiyama:2003cu,Lin:2024gcm}. Masses of hidden charmed tetraquarks consisting of a diquark and an antidiquark both with spin-$0$ are presented in Table~\ref{tab:hc00}, masses of hidden charmed tetraquarks consisting of a diquark and an antidiquark with one spin-$0$ and one spin-$1$ are presented in Table~\ref{tab:hc10}, and masses of hidden charmed tetraquarks consisting of a diquark and an antidiquark both with spin-$1$ are presented in Table~\ref{tab:hc11}.

\begin{table*}[htb]
\caption{\label{tab:hc00}Mass spectrum (in MeV) of hidden charmed tetraquarks consisting of a diquark and an antidiquark both with spin-$0$ in $\mathbf{\bar{3}}_c\otimes\mathbf{3}_c$ color configuration.}
\begin{ruledtabular}
\begin{tabular}{cccccccc}
$n|[s_d,s_{\bar d}]_S,L\rangle_{J}$      & $J^{PC}$ & Set I     & Set II    & Set III   & Set IV    & Ref.~\cite{Anwar:2018sol}\\
\colrule
$1|[0,0]_0,0\rangle_0$  & $0^{++}$ & $3832.43$ & $3832.13$ & $3832.29$ & $3832.12$ & $3577$ \\
$1|[0,0]_0,1\rangle_1$  & $1^{--}$ & $4219.64$ & $4228.37$ & $4221.94$ & $4229.92$ & $3960$ \\
$2|[0,0]_0,0\rangle_0$  & $0^{++}$ & $4384.96$ & $4389.81$ & $4395.08$ & $4400.06$ & $4111$ \\
$2|[0,0]_0,1\rangle_1$  & $1^{--}$ & $4647.70$ & $4654.99$ & $4656.09$ & $4662.86$ & $4353$
\end{tabular}
\end{ruledtabular}
\end{table*}

\begin{table*}[htb]
\caption{\label{tab:hc10}Mass spectrum (in MeV) of hidden charmed tetraquarks consisting of a diquark and an antidiquark with one spin-$0$ and one spin-$1$ in $\mathbf{\bar{3}}_c\otimes\mathbf{3}_c$ color configuration.}
\begin{ruledtabular}
\begin{tabular}{cccccccc}
$n|[s_d,s_{\bar d}]_S,L\rangle_{J}$ & $J^{PC}$ & Set I & Set II & Set III & Set IV   & Ref.~\cite{Anwar:2018sol}\\
\colrule
$1|[1/0,0/1]_1,0\rangle_{1}$   & $1^{++/-}$ & $3871.60$ & $3871.64$ & $3871.50$ & $3871.68$ & $3872$ \\
$1|[1/0,0/1]_1,1\rangle_{0}$   & $0^{--/+}$ & $4238.14$ & $4243.64$ & $4236.86$ & $4241.39$ & $4253$ \\
$1|[1/0,0/1]_1,1\rangle_{1}$   & $1^{--/+}$ & $4248.35$ & $4255.68$ & $4248.82$ & $4255.29$ & $4253$ \\
$1|[1/0,0/1]_1,1\rangle_{2}$   & $2^{--/+}$ & $4268.77$ & $4279.76$ & $4272.75$ & $4283.09$ &   -     \\
$2|[1/0,0/1]_1,0\rangle_{1}$   & $1^{++/-}$ & $4423.20$ & $4428.47$ & $4433.32$ & $4438.74$ & $4402$ \\
$2|[1/0,0/1]_1,1\rangle_{0}$   & $0^{--/+}$ & $4663.88$ & $4668.51$ & $4669.88$ & $4673.95$ & $4642$ \\
$2|[1/0,0/1]_1,1\rangle_{1}$   & $1^{--/+}$ & $4674.69$ & $4680.87$ & $4681.85$ & $4687.50$ & $4642$ \\
$2|[1/0,0/1]_1,1\rangle_{2}$   & $2^{--/+}$ & $4696.30$ & $4705.60$ & $4705.79$ & $4714.60$ &   -
\end{tabular}
\end{ruledtabular}
\end{table*}

\begin{table*}[htb]
\caption{\label{tab:hc11}
Mass spectrum (in MeV) of hidden charmed tetraquarks consisting of a diquark and an antidiquark both with spin-$1$ in $\mathbf{\bar{3}}_c\otimes\mathbf{3}_c$ color configuration.}
\begin{ruledtabular}
\begin{tabular}{ccccccc}
$n|[s_d,s_{\bar d}]_S,L\rangle_{J}$ & $J^{PC}$ & Set I & Set II & Set III & Set IV   & Ref.~\cite{Anwar:2018sol}\\
\colrule
$1|[1,1]_0,0\rangle_{0}$   & $0^{++}$ & $3762.65$ & $3760.47$ & $3769.81$ & $3768.55$ & $3641$ \\
$1|[1,1]_1,0\rangle_{1}$   & $1^{+-}$ & $3836.69$ & $3835.79$ & $3840.24$ & $3839.88$ & $4047$ \\
$1|[1,1]_2,0\rangle_{2}$   & $2^{++}$ & $3984.78$ & $3986.44$ & $3981.10$ & $3982.54$ &   -     \\
$1|[1,1]_0,1\rangle_{1}$   & $1^{--}$ & $4250.16$ & $4256.57$ & $4248.70$ & $4253.29$ & $4545$ \\
$1|[1,1]_1,1\rangle_{0}$   & $0^{-+}$ & $4231.70$ & $4235.25$ & $4227.32$ & $4229.73$ & $4567$ \\
$1|[1,1]_1,1\rangle_{1}$   & $1^{-+}$ & $4279.19$ & $4287.89$ & $4280.40$ & $4288.28$ &   -     \\
$1|[1,1]_1,1\rangle_{2}$   & $2^{-+}$ & $4284.59$ & $4295.62$ & $4287.76$ & $4298.10$ &   -     \\
$1|[1,1]_2,1\rangle_{1}$   & $1^{--}$ & $4276.30$ & $4281.85$ & $4274.55$ & $4279.97$ & $4570$ \\
$1|[1,1]_2,1\rangle_{2}$   & $2^{--}$ & $4325.24$ & $4334.72$ & $4327.92$ & $4336.65$ &   -     \\
$1|[1,1]_2,1\rangle_{3}$   & $3^{--}$ & $4333.34$ & $4346.31$ & $4338.96$ & $4351.37$ &   -     \\
$2|[1,1]_0,0\rangle_{0}$   & $0^{++}$ & $4405.17$ & $4411.60$ & $4419.46$ & $4426.10$ & $4482$ \\
$2|[1,1]_1,0\rangle_{1}$   & $1^{+-}$ & $4433.28$ & $4439.34$ & $4445.50$ & $4451.75$ & $4624$ \\
$2|[1,1]_2,0\rangle_{2}$   & $2^{++}$ & $4489.52$ & $4494.84$ & $4497.58$ & $4503.04$ &   -     \\
$2|[1,1]_0,1\rangle_{1}$   & $1^{--}$ & $4680.59$ & $4707.69$ & $4708.19$ & $4714.64$ & $4929$ \\
$2|[1,1]_1,1\rangle_{0}$   & $0^{-+}$ & $4662.41$ & $4665.67$ & $4666.74$ & $4669.41$ & $4947$ \\
$2|[1,1]_1,1\rangle_{1}$   & $1^{-+}$ & $4707.80$ & $4715.43$ & $4715.79$ & $4722.90$ &   -     \\
$2|[1,1]_1,1\rangle_{2}$   & $2^{-+}$ & $4715.47$ & $4725.09$ & $4724.76$ & $4733.90$ &   -     \\
$2|[1,1]_2,1\rangle_{1}$   & $1^{--}$ & $4701.24$ & $4684.60$ & $4685.59$ & $4688.47$ & $4949$ \\
$2|[1,1]_2,1\rangle_{2}$   & $2^{--}$ & $4744.90$ & $4752.79$ & $4753.13$ & $4760.50$ &   -    \\
$2|[1,1]_2,1\rangle_{3}$   & $3^{--}$ & $4756.40$ & $4767.28$ & $4766.59$ & $4776.99$ &   -
\end{tabular}
\end{ruledtabular}
\end{table*}

In these tables, the first column lists the tetraquark states in $LS$ coupling scheme from $1S$- to $2P$-wave excitations. The spin-parity quantum numbers $J^{PC}$ are given in the second column. The obtained mass spectra with the parameters sets I-IV of Ref.~\cite{Lin:2024gcm} are presented in the third to sixth columns. The masses of hidden charmed tetraquarks obtained with four sets of parameters are very close.

As a comparison, the results obtained in the relativized diquark-antidiquark model~\cite{Anwar:2018sol} are also listed in the last column in these tables. The lowest hidden charmed tetraquark is the $0^{++}$ one with mass $\sim 3832$ MeV, which is about $250$ MeV higher than the predicted one in Ref.~\cite{Anwar:2018sol}, while masses of hidden charmed tetraquarks consisting of a diquark and an antidiquark with one spin-$0$ and one spin-$1$ are close to the predicted ones in that method. For the states consisting of a diquark and an antidiquark both with spin-$1$, our results cover a narrower mass range.

From Tables~\ref{tab:hc00}-\ref{tab:hc11}, we present explicitly some low excitations of hidden charmed tetraquarks with different $J^{PC}$ in Table~\ref{jpc} for later convenience.

\begin{table*}[htb]
\caption{\label{jpc}Low excitations of hidden charmed tetraquarks with different $J^{PC}$.}
\begin{ruledtabular}
\begin{tabular}{ccccccc}
$J^{PC}$ & $n=1$ & $n=1$ & $n=2$ & $n=2$ & Observed states \\
\colrule
$0^{++}$ & $1|[1,1]_0,0\rangle_0(\sim 3770)$   & $1|[0,0]_0,0\rangle_0(\sim 3830)$ & $2|[0,0]_0,0\rangle_0(\sim 4390)$ & $2|[1,1]_0,0\rangle_0(\sim 4410)$ & $X^*(3860)$, X(4250), X(4350)  \\
$1^{+-}$ & $1|[1,1]_1,0\rangle_{1}(\sim 3840)$   & $1|[1/0,0/1]_1,0\rangle_{1}(\sim 3870)$  & $2|[1/0,0/1]_1,0\rangle_{1}(\sim 4430)$  & $2|[1,1]_1,0\rangle_{1}(\sim 4440)$ & $Z_c(3900)$,~X(3940),~$Z_c(4430)$  \\
$1^{++}$ & $1|[1/0,0/1]_1,0\rangle_{1}(\sim 3870)$ & -  & $2|[1/0,0/1]_1,0\rangle_{1}(\sim 4430)$  & - & X(3872) \\
$1^{--}$ & $1|[0,0]_0,1\rangle_1(\sim 4220)$   & $1|[1/0,0/1]_1,1\rangle_{1}(\sim 4250)$ & $2|[0,0]_0,1\rangle_0(\sim 4650)$ & $2|[1/0,0/1]_1,1\rangle_{1}(\sim 4680)$  & $Y(4230)$,~$X(Y)(4660)$  \\
$0^{-+}$ & $1|[1,1]_1,1\rangle_{0}(\sim 4230)$ & $1|[1/0,0/1]_1,1\rangle_{0}(\sim 4240)$ & $2|[1,1]_1,1\rangle_{0}(\sim 4660)$ & $2|[1/0,0/1]_1,1\rangle_{0} (\sim 4670)$ & X(4250),~$X(4630)$ \\
$0^{--}$  & -   & $1|[1/0,0/1]_1,1\rangle_{0}(\sim 4240)$ & - & $2|[1/0,0/1]_1,1\rangle_{0}(\sim 4670)$ & X(4240)  \\
$1^{-+}$ & $1|[1/0,0/1]_1,1\rangle_{1}(\sim 4250)$   & $1|[1,1]_1,1\rangle_{1}(\sim 4280)$ & $2|[1/0,0/1]_1,1\rangle_{1}(\sim 4680)$ & $2|[1,1]_1,1\rangle_{1}(\sim 4710)$ & X(4250), X(4630)
\end{tabular}
\end{ruledtabular}
\end{table*}
In Table~\ref{jpc}, the second to the fourth columns indicate the theoretical predictions, and the last column presents possible assignments of observed states.  For tetraquarks without radial excitations ($n=1$), the lowest hidden charmed tetraquark has mass $\sim 3770$ MeV with $J^{PC}=0^{++}$. This state consists of a diquark and an antidiquark both with spin-$1$. There is another hidden charmed $J^{PC}=0^{++}$ tetraquark nearby with mass $\sim 3830$ MeV, which has a diquark and an antidiquark both with spin-$0$. Their different inner structures may be distinguished from their hadronic decays. The mass of hidden-charmed tetraquark becomes higher in the $1^{+-},~1^{++},~1^{--},~0^{-+},~0^{--},~1^{-+}$, ... sequence. For tetraquarks with radial excitations ($n=2$), they have masses greater than $4300$ MeV.

So far, many XYZ exotics have been observed. Some XYZ exotics have been suggested as the normal $c\bar c$ charmoniums (see Table~\ref{candidates}) though there are controversial opinions. Based on the quantum numbers $J^{PC}$ and masses (uncertainties considered) from Table~\ref{candidates}, the following assignments of the observed XYZ exotics are natural suggestions.

$X^*(3860)$ was first observed by the Belle collaboration in the process $e^+e^-\to J/\Psi D\bar D$~\cite{Belle:2017egg}, but not seen by the LHCb collaboration in the decay $B^+\to D^+D^-K^+$~\cite{LHCb:2020pxc}. It could be assigned as the hidden charmed $0^{++}$ tetraquark composed of a diquark and an antidiquark both with spin-$1$. $Z_c(3900)$ and $X(3940)$ are possibly $1^{+-}$ hidden-charmed tetraquarks without radial excitation between the diquark and the antidiquark when theoretical uncertainties are taken into account. $X(3872)$ is fixed as the $1^{++}$ hidden charmed tetraquark consisting of a diquark and an antidiquark with one spin-$0$ and one spin-$1$. The quantum numbers $J$ and $P$ of $X(4250)$ have not been determined. For its large mass uncertainty in experiment, $X(4250)$ may be a $0^{-+}$ or $1^{-+}$ hidden charmed tetraquark consisting of a diquark and an antidiquark with one spin-$0$ and one spin-$1$ without radial excitation. If $X(4250)$ has mass around $4390$ MeV, it is possibly a hidden charmed $0^{++}$ tetraquark consisting of a diquark and an antidiquark both with spin-$0$ with radial excitation. $X(4240)$ may be a $0^{--}$ hidden charmed tetraquark consisting of a diquark and an antidiquark with one spin-$0$ and one spin-$1$.

$X(4350)$ may be a $0^{++}$ hidden charmed tetraquark with radial excitation. $Z_c(4430)$ may be a hidden charmed $1^{+-}$ tetraquark with radial excitation. The quantum numbers $J$ and $P$ of $X(4630)$ have not been determined yet, and there is also a large mass uncertainty for this state. $X(4630)$ may be a $0^{-+}$ or $1^{-+}$ hidden charmed tetraquark with radial excitation. $X(Y)(4660)$ may be the $1^{--}$ hidden charmed tetraquarks with radial excitation.

$Y(4260)$ was first observed by the BaBar collaboration in the $\pi^+\pi^-J/\Psi$ mass spectrum~\cite{BaBar:2005hhc}, and late experiment for $e^+e^-\to \pi^+\pi^-J/\Psi$ by the BESIII collaboration in a higher-statistics analysis~\cite{BESIII:2016bnd} indicates that $Y(4260)$ may consist of two resonances $Y(4230)$ and $Y(4320)$. $Y(4230)$ is possibly a $1^{--}$ hidden charmed tetraquark without radial excitation. $Y(4390)$ was observed in the processes $e^+e^-\to \pi^+\pi^-\Psi(3686)$ by the BESIII collaboration~\cite{BESIII:2017tqk}, and whether it is the $Y(4360)$ is unknown. $Y(4008)$ was first observed by the Belle Collaboration~\cite{Belle:2007dxy}, but was not confirmed by subsequent experiments by the BaBar and BESIII collaborations in the same production process. $Y(4008)$ was not thought as a genuine resonance, but a peak generated by the $\Psi(4040)$ with $D^*D$ loops with $\pi^+\pi^-J/\Psi$ in the final state~\cite{Piotrowska:2018rzl}. From the numerical results, there exists no $1^{--}$ hidden charmed vector tetraquark located between $4300$ and $4600$ MeV, or lower than $4200$ MeV. So it seems impossible for $Y(4008)$ or $Y(4390)$ to be the $1^{--}$ hidden charmed tetraquark. Of course, more decay and production features are required for the final identification of these exotics.

In order to find the spectra pattern of the hidden charmed tetraquarks, we have presented explicitly the mass splittings among different multiplets in Tables~\ref{tab:ms00}-\ref{tab:ms11}. The spin-weighted average mass of each multiplet is given by
\begin{equation}
\overline{M_{nL}}=\frac{\sum_J{(2J+1)M_{nL_{J}}}}{\sum_J{(2J+1)}}.
\end{equation}
For the mass splittings among different charmoniums in the last column in Tables~\ref{tab:ms00}-\ref{tab:ms11}, the masses of relevant charmoniums are listed in Table~\ref{tab:mcha}, where the masses of well established $1S$-, $1P$- and $2S$-wave charmoniums are taken from experimental data~\cite{ParticleDataGroup:2024cfk} while those of $2P$-wave charmoniums are extracted from Ref.~\cite{Barnes:2005pb}.

\begin{table*}[htb]
\caption{Mass splittings (in MeV) of radially and orbitally excited $[cq][\bar c\bar q]$ consisting of a diquark and an antidiquark both with spin-$0$, where the parameters are chosen as Sets I-IV.}
\label{tab:ms00}
\begin{ruledtabular}
\begin{tabular}{ccccccccccc}
 & Set I & Set II & Set III & Set IV  &  Charmonium~\cite{Barnes:2005pb,ParticleDataGroup:2024cfk} \\
\colrule
$1P-1S$ & $387.21$ & $396.24$ & $389.65$ & $397.80$ & $456.7$\\
$2S-1S$ & $552.53$ & $557.68$ & $562.79$ & $567.94$ & $605.4$\\
$2P-1P$ & $428.06$ & $426.62$ & $434.15$ & $432.94$ & $415.4$\\
$2P-2S$ & $262.74$ & $265.18$ & $261.01$ & $262.80$ & $266.8$
\end{tabular}
\end{ruledtabular}
\end{table*}

\begin{table*}[htb]
\caption{Mass splittings (in MeV) of radially and orbitally $[cq][\bar c\bar q]$ consisting of a diquark and an antidiquark with one spin-$0$ and one spin-$1$, where the parameters are chosen as Sets I-IV.}
\label{tab:ms10}
\begin{ruledtabular}
\begin{tabular}{ccccccccccc}
 & Set I & Set II & Set III & Set IV  &  Charmonium~\cite{Barnes:2005pb,ParticleDataGroup:2024cfk}\\
\colrule
$1P-1S$ & $386.96$ & $396.08$ & $389.29$ & $397.51$ & $456.7$\\
$2S-1S$ & $551.60$ & $556.83$ & $561.82$ & $567.06$ & $605.4$\\
$2P-1P$ & $426.93$ & $425.52$ & $433.03$ & $431.86$ & $415.4$\\
$2P-2S$ & $262.29$ & $264.77$ & $260.50$ & $262.31$ & $266.8$
\end{tabular}
\end{ruledtabular}
\end{table*}

\begin{table*}[htb]
\caption{Mass splittings (in MeV) of radially and orbitally $[cq][\bar c\bar q]$ consisting of a diquark and an antidiquark both with spin-$1$, where the parameters are chosen as Sets I-IV.}
\label{tab:ms11}
\begin{ruledtabular}
\begin{tabular}{ccccccccccc}
 & Set I & Set II & Set III & Set IV  &  Charmonium~\cite{Barnes:2005pb,ParticleDataGroup:2024cfk}\\
\colrule
$1P-1S$ & $386.72$ & $395.92$ & $388.94$ & $397.22$ & $456.7$\\
$2S-1S$ & $550.67$ & $555.98$ & $560.87$ & $566.18$ & $605.4$\\
$2P-1P$ & $425.81$ & $424.41$ & $431.91$ & $430.77$ & $415.4$\\
$2P-2S$ & $261.86$ & $264.36$ & $259.98$ & $261.80$ & $266.8$
\end{tabular}
\end{ruledtabular}
\end{table*}

\begin{table}[htb]
\caption{
Mass spectrum (in MeV) of charmonium~\cite{Barnes:2005pb,ParticleDataGroup:2024cfk}.}
\label{tab:mcha}
\begin{ruledtabular}
\begin{tabular}{ccccccccccc}
$n^{2S+1}L_J$ & $J^{PC}$ & Name & Mass~\cite{Barnes:2005pb,ParticleDataGroup:2024cfk} \\
\colrule
$1^1S_0$      & $0^{-+}$ & $\eta_c(1S)$     & $2983.9\pm0.4$                                                      \\
$1^3S_1$      & $1^{--}$ &  $J/\psi(1S)$    & $3096.900\pm0.006$                                                      \\
$1^1P_1$      & $1^{+-}$ &  $h_c(1P)$       & $3525.37\pm0.14$                                                     \\
$1^3P_0$      & $0^{++}$ &  $\chi_{c0}(1P)$    & $3414.71\pm0.30$                                                     \\
$1^3P_1$      & $1^{++}$ &  $\chi_{c1}(1P)$    & $3510.67\pm0.05$                                                     \\
$1^3P_2$      & $2^{++}$ &  $\chi_{c2}(1P)$    & $3556.17\pm0.07$                                                     \\
$2^1S_0$      & $0^{-+}$ &  $\eta_c(2S)$    & $3637.7\pm1.1$                                                      \\
$2^3S_1$      & $1^{--}$ &  $\psi(2S)$    & $3686.10\pm0.06$\\
$2^1P_1$      & $1^{+-}$ &  $h_c(2P)$       & $3934$                                                     \\
$2^3P_0$      & $0^{++}$ &  $\chi_{c0}(2P)$    & $3852$                                                     \\
$2^3P_1$      & $1^{++}$ &  $\chi_{c1}(2P)$    & $3925$                                                     \\
$2^3P_2$      & $2^{++}$ &  $\chi_{c2}(2P)$    & $3972$
\end{tabular}
\end{ruledtabular}
\end{table}

As shown in Tables~\ref{tab:ms00}-\ref{tab:ms11}, the corresponding mass splittings of three kinds of hidden charmed tetraquarks are very close. The mass splittings of $1P-1S$ and $2S-1S$ are about $390-400$ and $550-570$ MeV, respectively, which are about $50$ and $40$ MeV smaller than those of charmoniums. However, the mass splittings of $2P-1P$ and $2P-2S$ of hidden charmed tetraquarks are close to those of charmonium, which are about $250-260$ MeV and $420-430$ MeV, respectively. The feature for mass splittings exhibits a different spin-dependent contribution to the radial excitation in tetraquarks.

\section{summary}\label{summary}
In this work, a hidden charmed tetraquarks without strange or antistrange quark is assumed consisting of a pair of $cq$ diquark and $\bar c\bar q$ antidiquark. The interactions in tetraquark are modulated as two kinds of interactions: the first type is the interaction between a quark and another quark, or the interaction between an antiquark and another antiquark; the second type is the interaction between a diquark and an antiquark. For the practical calculation, the modified $AL$-type quark-quark and quark-antiquark potentials proposed by Semay and Silvestre-Brac are employed. To describe the interaction within a tetraquark in a better way, the parameters of the potential between the diquark and the antidiquark are refitted with the experimental mass of $X(3872)$ and $T_{cc}(3875)^+$.

In terms of the Semay-Silvestre-Brac potentials, the mass of the $cq$ diquark or $\bar c\bar q$ antidiquark with spin-$0$ is predicted with $\sim 2175$ MeV, and the mass of that with spin-$1$ is predicted with $\sim 2220$ MeV. The mass splitting between the scalar and vector diquark is $\sim 45$ MeV. The predicted mass of the $cq$ diquark with spin-$1$ is about $\sim 300$ MeV greater than the predicted one from QCD sum rule or phenomenological analysis.

Semay-Silvestre-Brac potentials are also employed to describe the interactions between the diquark and the antidiquark. When the parameters $\alpha$ and $\lambda$ are fixed by the two ground state tetraquark candidates $X(3872)$ and $T_{cc}(3875)$, the masses of hidden charmed tetraquarks from $1S$ to $2P$ excitations are calculated. The lowest hidden charmed tetraquark is predicted with mass $\sim 3770$ MeV and $J^{PC}=0^{++}$. The mass of hidden charmed tetraquark becomes higher in the $1^{+-},~1^{++},~1^{--},~0^{-+},~0^{--},~1^{-+}$, ... sequence, where the tetraquarks with exotic $J^{PC}=0^{--},~1^{-+}$ have greater masses. The hidden charmed tetraquark with radial excitation has mass greater than $4300$ MeV.

Based on the numerical results of the tetraquarks, some observed XYZ exotics have been analyzed and their possible $J^{PC}$ and inner structures are suggested.
$X^*(3860)$ could be assigned as the $0^{++}$ hidden charmed tetraquark consisting of a diquark and an antidiquark both with spin-$1$. $Z_c(3900)$ and $X(3940)$ are possibly the $1^{+-}$ hidden charmed tetraquarks without radial excitation. $X(3872)$ is fixed as a $1^{++}$ hidden charmed tetraquark consisting of a diquark and an antidiquark with one spin-$0$ and one spin-$1$. $X(4250)$ may be a $0^{-+}$ or $1^{-+}$ hidden charmed tetraquark without radial excitation, or a $0^{++}$ hidden charmed tetraquark with radial excitation. $X(4240)$ may be a $0^{--}$ hidden charmed tetraquark consisting of a diquark and an antidiquark with one spin-$0$ and one spin-$1$. $Y(4008)$ or $Y(4390)$ seems impossible to be the $1^{--}$ hidden charmed tetraquark.

With radial excitations, $X(4350)$ may be a $0^{++}$ hidden charmed tetraquark. $Z_c(4430)$ may be a $1^{+-}$ hidden charmed tetraquark. $X(4630)$ may be a $0^{-+}$ or $1^{-+}$ hidden charmed tetraquark. $X(Y)(4660)$ may be the $1^{--}$ hidden charmed tetraquarks. Obviously, these tentative assignments are based only on analyses of $J^{PC}$ quantum numbers and masses, more decay and production features are required for their final identification.

The masses of hidden charmed tetraquarks with a strange or antistrange quark have not been calculated, and few relevant exotics have not been analyzed because of the shortage of confirmed experimental data. There are some uncertainties in our calculation. First, the mixing between normal mesons and tetraquarks has not been taken into account explicitly, but was taken into account through the variation of parameters instead. The fixed parameters may vary with the components of the benchmark $X(3872)$ and $T_{cc}(3875)$, which will bring uncertainties of several tens of MeV to the predicted mass spectra. According to Ref.~\cite{Lin:2024olg}, boson-exchange interactions between the diquark and the antidiquark may make the predicted mass spectra lower than the predicted ones without boson-exchange interactions. So far, some charmonium-like XYZ exotics have been suggested as the normal $c\bar c$ charmoniums though these assignments are not definite. In order to identify a normal charmonium or charmonium-like meson, it is important to take into account the mixing between the normal charmonium and the hidden-flavor charmonium-like exotics at the same time. Only when the properties of normal charmonium and hidden-flavor charmonium-like exotics are clear both in theories and in experiments, the identification could be confirmed.

\begin{acknowledgments}
This work is supported by National Natural Science Foundation of China under Grant No. 11975146. Ailin Zhang is grateful for the useful discussions with Changzheng Yuan.
\end{acknowledgments}


%

\end{document}